\documentclass{aa}
\include{psfig}
\def\etal{{\it et al.\ }}

\def\kms{\ {\rm km\,s^{-1}}}
\usepackage{graphics}
\begin{document}
\thesaurus{12(11.03.5; 11.04.1; 11.05.1; 11.11.1 - 12.03.3)}
\title{Dynamics of the X-ray clusters Abell 222, Abell 223 and Abell 520
\thanks{based on observations made at ESO-La Silla (Chile), at the Canada 
France Hawaii Telescope and at the Pic du Midi Observatory (France).}}
\author{D.~Proust\inst{1} \and H. Cuevas\inst{2} \and H.V. Capelato \inst{3} 
\and L. Sodr\'e Jr.\inst{2} \and B. Tom\'e Lehodey \inst{3}
\and O. Le F\`evre \inst{4} \and A. Mazure \inst{4}}
\offprints{D.~Proust}
\institute{DAEC, CNRS, Observatoire de Paris-Meudon, 92195 Meudon Cedex,
France.
\and
Departamento de Astronomia IAG/USP, Av. Miguel Stefano 4200, 04301-904, 
S\~ao Paulo, Brazil
\and
Divis\~ao de Astrofisica INPE/MCT, 12225-010, S\~ao Jos\'e dos Campos, Brazil
\and
Laboratoire d'Astronomie Spatiale, Les Trois Lucs, 13012 Marseille, France}

\date{Received date; accepted date}
\titlerunning{X-ray clusters of galaxies}
\maketitle

\begin{abstract}

We present the results of a dynamical analysis of three rich, X-ray
luminous galaxy clusters, Abell~222, Abell~223 and Abell~520, that
are at intermediate redshifts. Our study is based on
radial velocities for 71 cluster members, respectively 30 for A222, 20
for A223 and 21 for A520, measured from spectra obtained at the
Canada-France-Hawaii Telescope, the European Southern Observatory, and
the Pic du Midi Observatory, and supplemented with radial velocities
from the literature.  A222 galaxies have slightly higher velocities
than those of A223, with bi-weighted mean velocity of $V_{bi} = 64242
\pm 194~\kms$ for A222, and of $V_{bi} = 63197 \pm 266~\kms$ for A223.
The velocity dispersions of the two clusters are about the same:
$\sigma_{bi} = 1013 \pm 150~\kms$ and $\sigma_{bi} = 1058 \pm 160~\kms$
for A222 and A223, respectively. For A520 we obtain $V_{bi} = 60127 \pm
284~\kms$ with $\sigma_{bi} = 1250 \pm 189~\kms$. We also give mass and
mass-luminosities ratio estimates for each cluster separately.  We
argue that these clusters are presently undergoing strong dynamical
evolution and that A222 and A223 will probably merge in the future.  We
have applied a Principal Component Analysis to a sample of 51  CFHT
spectra to produce a spectral classification for these galaxies. 
This classification has allowed us to show that the 
morphological and kinematical
segregations  were already established in these intermediate redshift
clusters.

\end{abstract}

\section{Introduction} 

Clusters of galaxies are the largest gravitationally bound systems,
with the following components contributing to their total mass:  dark
matter, which is the dominant component, the hot X-ray emitting gas,
that is the dominant baryonic component, and the stars and gas in galaxies.  A
valuable approach to determine the distribution of these components is
offered by studying the relation between the global cluster properties
which can be directly measured, such as the velocity
dispersion~$\sigma$, the total luminosity~$L$, effective radii~$R_{e}$,
and morphological type distribution.  Correlations between these
intrinsic parameters have been found for many galaxy clusters, e.g.
between richness and velocity dispersion (Danese, de Zotti \& di
Tullio 1980, Cole
1989), and between radius and luminosity (West, Oemler \& Deckel 1989).  
Evidence
pointing towards a close link between morphology and environment are
the morphological segregation (Dressler 1980) and the correlation
between type and velocity dispersion (Sodr\'e et al. 1989), which we
call kinematical segregation.  Knowing whether these phenomena are due
to initial conditions, environmental effects or both is one of the main
questions to be answered in the study of these structures.

Morphological types, however, can be expensive to obtain. An
interesting alternative is to use spectral classification to obtain
spectral types (Sodr\'e \& Cuevas 1994, 1997;
Folkes, Lahav \& Maddox 1996). This procedure is based on a principal
component analysis (PCA) of the spectra and allows to define a spectral
classification that presents some advantages over the usual
morphological classification: it provides quantitative, continuous and
well defined types, avoiding the ambiguities of the intrinsically more
qualitative and subjective morphological classification. This method
has also been applied to the study of the ESO-Sculptor Survey (Galaz \& de
Lapparent 1998) and to the  Las Campanas Redshift Survey (Bromley et
al. 1998), and it was found that PCA allows to classify galaxies in
an ordered and continuous spectral sequence, which is strongly
correlated with the morphological type.
 
In this paper we present an analysis of three medium distant X-ray
clusters, A222, A223 and A520, all of them belonging to the Butcher,
Oemler \& Wells (1983, hereafter BOW83) photometric sample. 
For these clusters,
very few redshifts exist, as found in the NED database\footnote{The
NASA/IPAC Extragalactic Database (NED) is operated by the Jet
Propulsion Laboratory, California Institute of Technology, under
contract with the National Aeronautics and Space Administration}. We
have obtained new spectra which have enabled us to perform a
preliminary study of their dynamical properties. The projected galactic
densities of the clusters were compared to the X-ray emission images in
order to seek for substructures inside these systems. From the observed
spectra we have proceeded the spectral classification of the galaxies,
allowing, for the first time, to study the morphological and
kinematical segregations present in these clusters.  In Section~2 we
describe the observations, instrumentation, data reduction techniques
and comparisons with previous measurements.  In Section~3, the spatial
distribution and kinematical properties are performed and discussed as
well as mass and mass-luminosities ratio estimates. In Section~4 we
analyze the spectral classification of the galaxies. Section~5 discuss
the morphological and kinematical segregations. Finally, in Section 6
we present a summary of our main results.

\section{Observations and Data Reductions}

The analysis of the dynamical state of the clusters 
discussed here is based on a
large set of velocities for the cluster galaxies. Multi-object
spectroscopy has been performed at CFHT in November 1993 and at the ESO
3.60m telescope in December 1995. The instrumentation used at CFHT was
the Multi Object Spectrograph (MOS) using the grism O300 with
a dispersion of 240~\AA/mm and the STIS CCD of 2048x2048 pixels of
21$\mu$m, leading to a dispersion of 5~\AA/pixel. The instrumentation
used at ESO was the ESO Faint Object Spectrograph and Camera (EFOSC) with
the grism O300 yielding a dispersion of 230\AA/mm and the TEK512 CCD
chip of 512x512 pixels of 27$\mu$m giving a resulting dispersion of
6.3~\AA/pixel. We completed the observations during an observing run at
the 2.0m Bernard Lyot telescope at Pic du Midi observatory in January
1997, using the ISARD spectrograph in its long-slit mode with a
dispersion of 233~\AA/mm and with the TEK chip of 1024x1024 pixels of
25$\mu$m, corresponding to 5.8~\AA/pixel. Typically, two exposures of
2700s each were taken for fields across the cluster. Wavelength
calibration was done using arc lamps before each exposure (Helium-Argon
at CFHT, Helium-Neon at ESO and Mercury-Neon at Pic du Midi lamps).
 
The data reduction was carried out with IRAF\footnote{IRAF is
distributed by the National Optical Astronomy Observatories, which are
operated by the Association of Universities for Research in Astronomy,
Inc., under cooperative agreement with the National Science
Foundation.} using the MULTIRED package (Le F\`evre \etal 1995).
The sky spectrum has been removed from the data in each slit
by using measurements at each side of the galaxy spectra.

Radial velocities have been determined using the cross-correlation
technique (Tonry and Davis 1979) implemented in the RVSAO package
(Kurtz \etal 1991, Mink \& Wyatt 1995) with radial velocity standards
obtained from observations of late-type stars and previously
well-studied galaxies.

From the total set of data we have retained 78
successful spectra of objects (28~for A222, 25~for A223 and 25~for
A520) with a signal-to-noise ratio high enough to derive the
measurement of the radial velocity with good confidence, as indicated
in Table~1 by the R value of Tonry and Davis (1979). Note that
since the templates used in the reduction were the same for all spectra,
the R values contained in Table 1 are proportional to the signal-to-noise
ratio of the spectra. Star contamination
was very low (only 1 of the selected targets turned out to be a star).
To help the reader to appreciate the kind of data discussed in this paper,
we present in Figure 1 two spectra, one with high signal-to-noise ratio
(R=6.59) and the other with low signal-to-noise ratio (R=3.96). 

\begin{figure}
\psfig{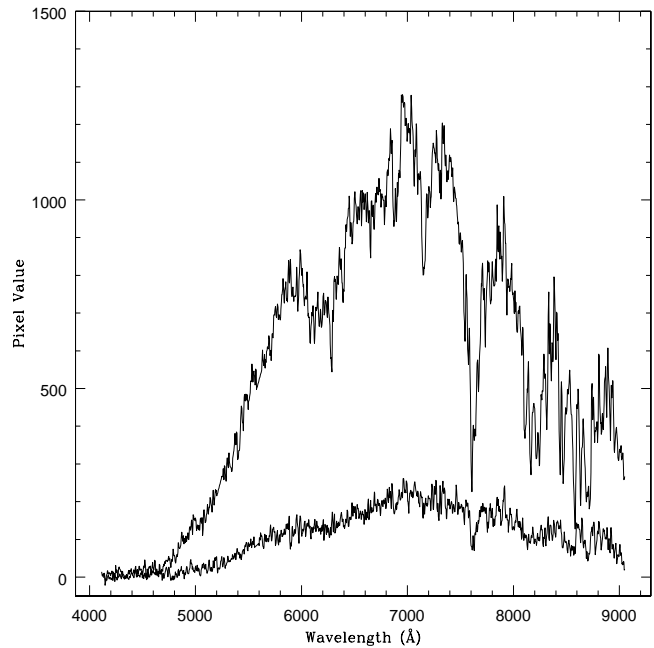}
\caption[]{ Examples of two CFHT spectra of A222: one of a
bright galaxy ($f_{57}=17.87$) with a high signal-to-noise ratio 
(R=6.59) and the other of a fainter galaxy ($f_{57}=18.98$) with
low signal-to-noise ratio (R=3.96).}
\end{figure}

Table~1 lists positions and heliocentric velocities for the
78~individual galaxies in the clusters. For each galaxy we give also
$f$ and $j$ band photometry from BOW83. The table is
completed with a few galaxies observed by Newberry, Kirshner \&
Boroson (1988) and
Sandage, Kristian \& Westphal (1976). The table entries are:

\begin{enumerate}
\item galaxy number

\item right ascension (hour, min, sec)

\item declination (degree, minute, second)

\item $f_{57}$ magnitude from Butcher \etal (1983).

\item $j-f_{57}$ color from Butcher \etal (1983).

\item heliocentric radial velocity with its error in $\kms$

\item R-value derived from Tonry \& Davis (1979).

\item instrumentation and notes,  {\bf c}: 3.60m CFHT telescope,
{\bf e}: 3.60m ESO telescope, {\bf l}: Newberry,  Kirshner \&
Boroson (1988),  {\bf p}: 2.0m BL telescope, {\bf s}: Sandage, 
Kristian \& Westphal (1976).

\end{enumerate}

\begin{table*}
\caption[]{Heliocentric redshifts for galaxies.}
\begin{flushleft}
\begin{tabular}{llllllll}
\hline
{\bf GALAXY} & {\bf R.A.} & {\bf DEC.} & {\bf $f_{57}$ mag} &
{\bf $j-f_{57}$ mag} & {\bf HEL. VEL.} & {\bf TDR} & {\bf N} \\
             &   (2000)   &   (2000)   &                    &
                     & $V {\pm {\Delta}V}$ & value &         \\ 
\hline
{\bf A222} &          &             &       &      &           &       &   \\
        & 01 37 10.00 & -12 59 56.7 &       &      & 63395 123 &  2.56 & c \\
        & 01 37 14.54 & -12 59 20.4 &       &      & 64999  76 &  4.13 & c \\
        & 01 37 19.75 & -13 00 42.3 & 18.81 & 1.66 & 64982  77 &  5.70 & c \\
        & 01 37 21.32 & -13 00 39.9 & 18.40 & 1.68 & 66371  64 &  4.26 & c \\
        & 01 37 21.34 & -12 59 22.9 & 19.99 & 1.62 & 63628  70 &  4.79 & e \\
        & 01 37 22.05 & -12 58 59.9 & 20.71 & 1.17 & 62646  68 &  2.85 & e \\
        & 01 37 22.72 & -13 00 23.0 & 18.33 & 1.61 & 63441  82 &  5.31 & c \\
        & 01 37 24.57 & -12 57 50.8 & 20.98 & 0.90 & 82493  85 &  2.32 & e \\
        & 01 37 25.22 & -13 00 21.7 & 20.80 & 1.49 & 66352  68 &  2.91 & e \\
        & 01 37 25.35 & -12 59 50.8 & 19.47 & 1.29 & 51476  43 &  3.33 & c \\
        & 01 37 26.45 & -12 59 58.5 & 17.91 & 1.83 & 64024  21 &  6.16 & c \\
        & 01 37 28.02 & -13 00 03.0 & 19.71 & 1.60 & 64178  40 &  9.97 & e \\
        & 01 37 28.96 & -12 59 26.4 & 19.09 & 0.94 & 54700  ?? &       & c \\
        &             &             &       &      & 54863  30 &       & l \\
        & 01 37 29.53 & -12 59 50.2 & 19.73 & 1.51 & 65510 104 &  3.62 & e \\
        & 01 37 29.70 & -13 00 26.8 & 20.29 & 1.69 & 63869 110 &  3.46 & c \\
        & 01 37 30.88 & -12 59 25.2 & 19.38 & 1.69 & 64859  20 & 10.42 & e \\
        &             &             &       &      & 64957  25 &  5.48 & c \\
        &             &             &       &      & 64847 210 &       & l \\
        & 01 37 31.64 & -12 58 52.4 & 19.52 & 1.69 & 65025  33 &  8.18 & e \\
        & 01 37 32.63 & -12 58 43.7 & 20.56 & 1.41 & 65274  74 &  4.98 & e \\
        & 01 37 32.74 & -12 59 19.0 & 20.35 & 1.62 & 64148  70 &  3.18 & c \\
        & 01 37 33.50 & -12 59 00.0 & 20.17 & 1.56 & 64448  28 &  9.46 & e \\
        & 01 37 33.49 & -12 59 28.0 & 18.69 & 1.63 & 65057 500 &       & s \\
        & 01 37 34.13 & -12 59 30.1 & 17.87 & 1.83 & 64115  37 &  6.59 & c \\
        & 01 37 34.48 & -12 59 46.3 & 18.98 & 1.79 & 65895  65 &  3.96 & c \\
        & 01 37 34.54 & -12 59 24.0 & 19.56 & 1.66 & 63857  90 &       & l \\
        & 01 37 34.61 & -12 58 41.1 & 19.13 & 1.68 & 62911  33 & 10.14 & e \\
        & 01 37 35.59 & -12 59 26.8 & 19.01 & 1.46 & 62974  46 &  5.28 & c \\
        &             &             &       &      & 62748 150 &       & l \\
        & 01 37 37.24 & -12 59 16.0 & 19.32 & 0.82 & 64277  60 &       & l \\
        & 01 37 38.67 & -12 59 24.4 & 20.49 & 1.55 & 62504 122 &  2.30 & c \\
        & 01 37 40.06 & -12 58 31.0 & 19.16 & 1.15 & 63887 180 &       & l \\
        & 01 37 40.18 & -12 59 56.6 & 19.92 & 1.62 & 63430  54 &  3.52 & c \\
        & 01 37 41.59 & -12 58 31.9 & 17.98 & 1.72 & 64181  67 &  3.68 & c \\
        & 01 37 42.60 & -13 00 37.7 & 19.69 & 1.59 & 64370 112 &  3.62 & c1\\
        & 01 37 43.00 & -12 57 44.0 & 18.31 & 1.82 & 64067  30 &       & l \\
{\bf A223}      &     &             &       &      &           &       &   \\ 
        & 01 37 32.31 & -12 52 43.7 &       &      & z=0.3063  &       & c2\\
        & 01 37 35.38 & -12 49 57.3 &       &      & 62304  22 &  6.53 & c \\
        & 01 37 36.08 & -12 53 42.6 &       &      & 62461  71 &  3.87 & c \\
        & 01 37 40.75 & -12 52 00.9 &       &      & 63782  72 &  5.46 & c \\
        & 01 37 41.62 & -12 52 15.4 &       &      & 62387  72 &  4.54 & c \\
        & 01 37 44.02 & -12 51 10.1 & 19.83 & 0.93 & 82918  36 &       & c3\\
        & 01 37 44.84 & -12 50 57.3 & 19.47 & 1.79 & 83377  40 &  6.20 & c \\
        & 01 37 44.86 & -12 44 48.5 & 19.91 & 1.73 & 82227  43 &  5.05 & c \\
        & 01 37 45.48 & -12 46 27.1 & 19.04 & 1.47 & 61485     &       & c4\\
        & 01 37 46.01 & -12 49 49.4 & 19.34 & 1.57 & 83432  57 &  5.23 & c \\
        & 01 37 49.59 & -12 51 37.6 & 18.88 & 1.63 & 64595  50 &  7.46 & c \\
        & 01 37 50.82 & -12 51 01.1 & 19.75 & 1.60 & 63738  98 &  3.88 & c \\
        & 01 37 51.16 & -12 47 33.7 & 19.31 & 0.67 & z=0.4726  &       & c5\\
        & 01 37 52.14 & -12 47 41.4 & 18.98 & 1.56 & 65350  89 &  3.61 & c \\
\hline
\end{tabular}
\end{flushleft}
\end{table*}
\begin{table*}
\begin{flushleft}
\begin{tabular}{llllllll}
\hline
{\bf GALAXY} & {\bf R.A.} & {\bf DEC.} & {\bf $f_{57}$ mag} &
{\bf $j-f_{57}$ mag} & {\bf HEL. VEL.} & {\bf TDR} & {\bf N} \\
             &   (2000)   &   (2000)   &                    &
                     & $V {\pm {\Delta}V}$ & value &         \\
\hline
        & 01 37 52.23 & -12 49 54.0 & 19.48 & 1.58 & 64440  98 &  5.18 & c \\
        & 01 37 53.04 & -12 50 15.3 & 19.41 & 1.63 &  star     &       & c \\
        & 01 37 54.96 & -12 49 55.0 & 19.43 & 1.57 & 63273  79 &  4.22 & c6\\
        & 01 37 55.35 & -12 51 31.1 & 19.39 & 1.57 & 63200  41 &  5.47 & c \\
        & 01 37 56.07 & -12 49 11.7 & 18.33 & 1.68 & 63716  89 &  2.96 & c \\
        & 01 37 56.12 & -12 48 17.0 & 18.37 & 1.68 & 62778 120 &       & l \\
        & 01 37 56.89 & -12 49 12.6 & 19.12 & 1.58 & 62341     &       & c7\\
        & 01 37 57.28 & -12 50 18.9 & 18.20 & 1.33 & 62473  11 &       & c8\\
        &             &             &       &      & 62208  90 &       & l \\
        & 01 37 57.18 & -12 48 51.0 & 18.66 & 0.87 & 62358 500 &       & s \\
        & 01 37 57.48 & -12 47 57.0 & 18.44 & 1.71 & 62868 120 &       & l \\
        & 01 37 58.14 & -12 50 55.5 & 19.31 & 0.99 & 63966  89 &  3.26 & c \\
        & 01 37 58.24 & -12 49 22.0 & 18.58 & 1.35 & 41732  90 &       & l \\
        & 01 37 59.90 & -12 47 34.6 & 20.56 & 1.33 & 64192  98 &  2.57 & c \\
        & 01 38 02.40 & -12 45 21.7 & 17.63 & 0.00 & 61532  39 &  3.43 & c \\
        &             &             &       &      & 61476  92 &  2.45 & p \\
        &             &             &       &      & 61758 500 &       & s \\ 
        & 01 38 04.12 & -12 44 50.4 & 17.73 & 0.00 & 15209  47 &  3.81 & p \\
{\bf A520} &          &             &       &      &           &       &   \\
        & 04 54 01.23 & +02 57 46.0 & 17.00 & 1.99 & 62172  37 &  4.71 & c \\
        & 04 54 05.12 & +02 53 30.0 & 17.72 & 1.22 & 60859 500 &       & s \\ 
        & 04 54 05.63 & +02 57 01.7 & 18.30 & 0.94 & 61145  75 &  3.51 & c \\
        & 04 54 05.98 & +02 55 55.4 & 19.47 & 1.97 & 60032  60 &  3.72 & c \\
        & 04 54 07.24 & +02 57 48.4 & 19.60 & 1.16 & 66926  65 &       & c9\\
        & 04 54 07.92 & +02 57 03.3 & 19.84 & 2.17 & 67830  78 &  2.93 & c \\
        & 04 54 09.51 & +02 58 18.1 & 19.34 & 1.89 & 59498  63 &  2.45 & c \\
        & 04 54 12.01 & +02 58 07.8 & 17.55 & 2.03 & 62256  35 &  6.32 & c \\
        & 04 54 13.14 & +02 57 33.8 & 17.17 & 1.96 & 60115  16 &  4.96 & c \\
        & 04 54 14.10 & +02 57 09.9 & 17.27 & 2.04 & 59506  46 &  4.87 & c \\
        &             &             &       &      & 59420  60 &       & l \\
        & 04 54 15.18 & +02 57 08.2 & 17.89 & 1.96 & 59224  92 &  4.52 & c \\
        & 04 54 16.14 & +02 56 42.7 & 17.87 & 2.01 & 58808  87 &  3.14 & c \\
        & 04 54 17.33 & +02 56 46.1 & 20.22 & 1.91 & 64056  81 &  3.07 & c \\
        & 04 54 19.27 & +02 58 26.3 & 18.46 & 1.93 & 60941  96 &  2.28 & c \\
        & 04 54 19.94 & +02 57 44.6 & 17.02 & 2.03 & 60381  53 &  3.91 & c \\
        & 04 54 20.30 & +02 55 37.0 & 17.08 & 1.94 & 58581  60 &       & l \\
        & 04 54 20.72 & +02 55 29.5 & 17.89 & 1.98 & 58482  97 &  2.34 & c \\
        & 04 54 21.41 & +02 56 47.6 & 18.96 & 2.04 & 60118  80 &  2.67 & c \\
        & 04 54 22.24 & +02 55 07.0 & 18.88 & 2.06 & 59150 300 &       & l \\
        & 04 54 23.23 & +02 56 40.8 & 18.86 & 1.95 & 73474  83 &  2.53 & c \\
        & 04 54 25.00 & +02 58 57.5 & 19.04 & 1.43 & 60370  71 &  3.47 & c \\
        & 04 54 25.52 & +02 59 37.5 & 20.85 & 1.84 & 60315  98 &  2.21 & c \\
        & 04 54 27.87 & +02 55 28.4 & 19.25 & 2.39 & 62563  91 &  2.86 & c \\
        & 04 54 28.95 & +02 56 46.6 &       &      & 69839  98 &  2.34 & c \\
        & 04 54 31.46 & +02 57 24.7 &       &      & z= 0.373  &       & c10\\
        & 04 54 32.74 & +02 54 48.3 &       &      & 60862  49 &  5.36 & c \\
        & 04 54 34.68 & +02 56 50.6 &       &      & 58629  57 &  2.56 & c \\
        &             &             &       &      & z= 0.334  &       & c11\\
\hline
\end{tabular}
\end{flushleft}
Notes.{\bf 1}: H${\beta}=~64420\kms$; {\bf 2}: measured on [OII],
H${\beta}$, [OIII] and $H{\alpha}$; {\bf 3}: measured on the strong
H${\alpha}$; {\bf 4}: measured on the strong H${\beta}$; {\bf 5}: measured
on [OII]; {\bf 6}: H${\beta}: 63329~\kms$;
{\bf 7} and {\bf 8}: measured on [OII], H${\beta}$ and [OIII];
{\bf 9}: measured on H${\alpha}$; {\bf 10}: measured on H and K lines;
{\bf 11}: second object on the line of sight.
\end{table*}

In order to test the external accuracy of our velocities, we compared our
redshift determinations ($V_{P}$) with data available in 
literature ($V_{L}$) for 5 galaxies observed in common (4 from
Newberry, Kirshner \& Boroson 1988, and 1 from Sandage, Kristian and
Westphal 1976). The mean value of $(V_P - V_L)$ is very small, 27 km s$^{-1}$,
and the null hypothesis that these two sets have the same variance but
significantly different means can be rejected at the 99\% level.

\section{Spatial distribution and kinematical properties}

\subsection{The binary cluster A222 + A223}
As already noticed by Sandage, Kristian \& Westphal 
(1976), these two neighboring
clusters have nearly the same redshift and probably constitute an
interacting system which is going to merge in the future.  Both are
dominated by a particularly bright cD galaxy.  They have a richness
class R=3 and are X-ray luminous with $L_{X}(7~{\rm keV})=~3.7 \pm
0.7~10^{44}~{\rm erg~s}^{-1}$ and $1.5 \pm 0.6~10^{44}~{\rm erg~s}^{-1}$ 
for A222 and A223, respectively
(Lea and Henry, 1988). The BOW83 sample covers only the central regions
of these two clusters and, in order to study the galaxy distribution in
these systems, as well as to estimate the projected density for the
galaxies in our sample (see below), we have built a more extensive,
although shallower, galaxy catalog, covering a region of 
$45^\prime \times 45^\prime$ 
centered on the median position of the two clusters.  This catalogue,
with 356 objects, was extracted from Digital Sky Survey (DSS) images,
using the software SExtractor ( Bertin \& Arnouts 1996).  It is more
than 90\% complete to BOW83 magnitudes $f_{57} \sim 19$.

Figure~2 displays the significance map for the projected densities of
galaxies in the region (cf. Biviano et al. 1996 for details on this type of
map), as derived from the DSS sample. The overall distribution of galaxies
is elongated along the direction defined by the two main clusters,
showing extension from both sides and suggesting that both clusters
belong to the same -- probably still collapsing -- structure.

\begin{figure}
\psfig{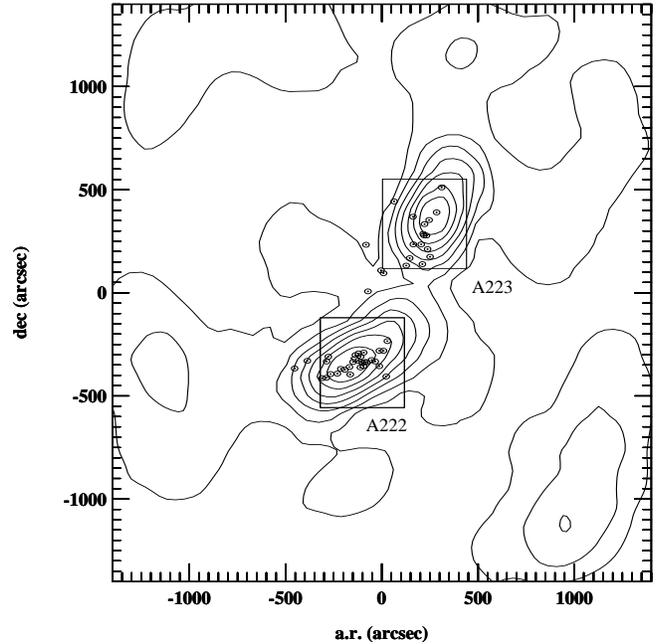}
\caption[]{Projected density map of the galaxies in
the field of A222 and A223 (centered at $\alpha_{2000} = 01^h 37^m 41^s$, 
$\delta_{2000} = -12^\circ 53^\prime 50^{\prime \prime}$). 
The galaxies were extracted from DSS images with the software
SExtractor (see text).  The circles with dots correspond to galaxies 
with measured radial velocities and belonging to the cluster. The 
squares display the areas covered by the BOW83 catalog.}
\end{figure}

In Figures 3a and 3b we display the isophotes of a wavelet
reconstruction (Ru\'e \& Bijaoui, 1997) of a ROSAT HRI X-ray image,
superposed on the significance maps of the projected density of
galaxies (dashed lines). In contrast with Figure 2 above, these maps
were constructed by taking galaxies from the deeper BOW83 catalog, which
is complete up to $f_{57} \sim 22$. This resulted in much more
features in the density maps than above, due to the introduction of the
faint galaxy population of the clusters. The X-ray emission roughly
follows the density contours of these maps. However, the apparent
regularity of the X-ray isophotes may be due to the smoothing effect
of the wavelet reconstruction, which favors larger scales against
the smallest ones.

These figures show that the general structure of both clusters, A222
and A223, is extremely complex, presenting several clumps of galaxies
in projection on their central regions, the reality of which is hard to
assert in the absence of much more radial velocity data than that
provided in this paper. Moreover, the fact that this complexity is not
seen in the brighter DSS projected density, indicates that the
projected clumps are mainly populated by faint galaxies.  The X-ray
emission is centered, for both clusters, in their main galaxy
concentrations, but this does not correspond to location of their
brightest members, as it is usually observed in nearby rich clusters of
galaxies. All these pieces of data support the view that we are facing
a dynamically unrelaxed, young system.

\begin{figure}
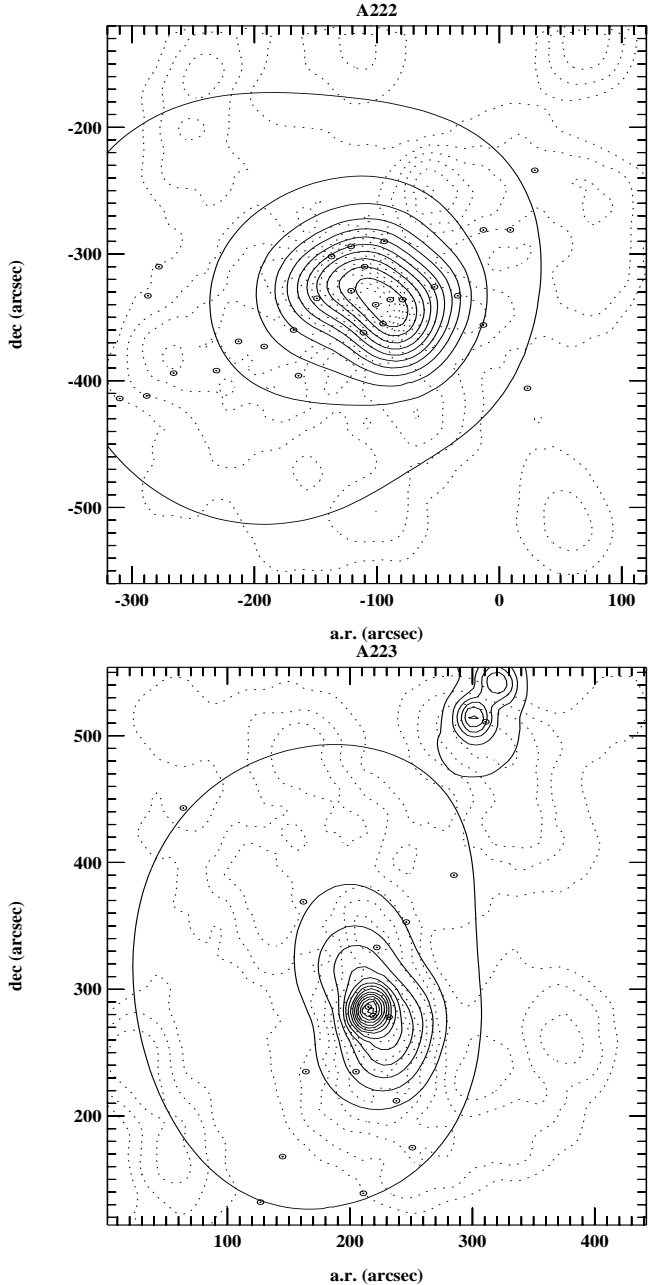

\psfig{figure=FIG3a.epsi,height=8.5cm,width=8.5cm,angle=-90}
\psfig{figure=FIG3b.epsi,height=8.5cm,width=8.5cm,angle=-90}
\caption[]{{\bf a.}X-ray isophotes from the ROSAT/HRI image of A222.
A mean background has been subtracted from the original image which 
was then reconstructed using wavelet transform and excluding the
smallest scales, due to intrinsic photon noise. The dashed lines
give the significance map of the galaxy projected densities from
the BOW83 catalog (only the highest density levels are displayed).
Other symbols have the same meaning as for Figure~2.
{\bf b.} Like Figure~3a but for A223.}
\end{figure}

For the more X-ray luminous cluster A223 (Figure 3b), we notice
the presence of an extended emission at North, centered near
the position of the brightest galaxy of the cluster (the Northeast
one). This emission is almost coincident with a projected substructure
of galaxies delineated by the isopleth curves, suggesting that this is
a real feature of the cluster.

We have used the ROSTAT statistical package (Beers et al. 1990; see also
Ribeiro et al. 1998 and references therein) to analyze the velocity
distributions obtained in this paper. ROSTAT provides several robust
estimators for the location, scale and shape of one-dimensional
data sets.  It includes a variety of normality tests as well as a
conservative unimodality test of the distribution (the Dip test, see
Hartigan \& Hartigan 1985). The shape estimators given by ROSTAT are
the Tail Index (TI) and the Asymmetry Index (AI) (for a thorough
discussion on these two indexes, see Bird \& Beers 1993). Since we
will be dealing with poor samples, we have restricted ourselves to the
use of the so called biweighted estimators of location and scale
(see Beers et al. 1990 for a definition), which generally
perform better in these cases. We notice, however, that the values of the
biweighted estimators obtained here differ negligeably from that
obtained using other commonly used estimators such as the 
conventional mean and dispersion obtained from recursive $3\sigma-$
clipping (Yahil \& Vidal, 1977), or median estimators.

Our radial velocity sample for the A222 and A223 system consists of 53
spectroscopically measured galaxies to which we added another 9 taken
from the literature (see Table 1).  Although by no means complete, this
sample is spatially reasonably well distributed to allow us some
preliminary analysis. Figure 4 shows the corresponding wedge velocity
diagram in right ascension and declination for A222 and A223.

\begin{figure*}
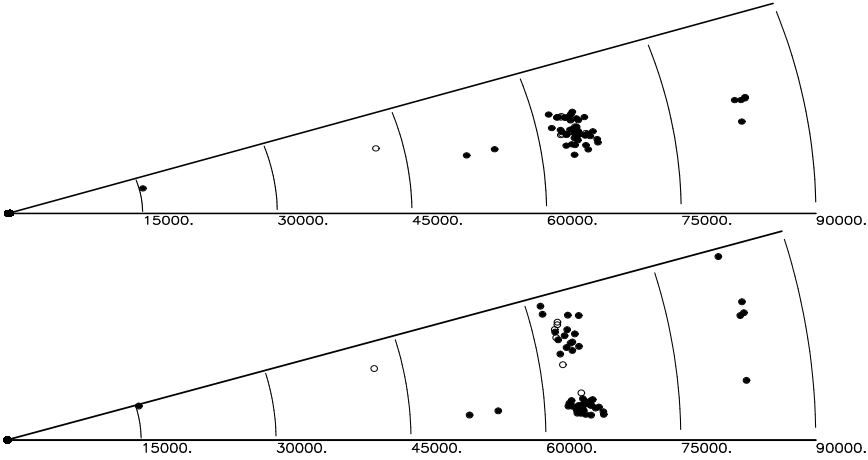

\psfig{figure=FIG4a.epsi,height=3cm,width=11.5cm,angle=-90}
\psfig{figure=FIG4b.epsi,height=3cm,width=11.5cm,angle=-90}
\caption{Wedge velocity diagram in right
ascension (up), and declination (down) for the measured galaxies in
A222 and A223. Filled symbols represent galaxies measured in the present work;
open symbols represent galaxies with velocities from the literature.}
\end{figure*}

After removing some obvious background and foreground galaxies -- also
confirmed by the recursive $3\sigma$ clipping -- we obtained a sample of 50
galaxies with measured velocities corresponding to the main peak seen
in the inset of the upper panel of Figure 5, which displays the radial
velocity distribution for the whole observed sample. The normality
tests provided by the ROSTAT package fail to reject a Gaussian parent
population for this sample. However, the Dip statistics has a value of
0.067, which is enough to reject the null hypothesis of unimodality at
significance levels better than 10\%. This is understandable, for this
sample refers to both components of the binary system A222 and A223.
Its mean velocity is $V_{bi} = 63833 \pm 165$ km s$^{-1}$, with
dispersion $\sigma_{bi} = 1157 \pm 119$ km s$^{-1}$. This places the
system at redshift $ z = 0.21292 $, slightly higher than the value
quoted by Strubble \& Rood (1987).

\begin{figure}
\psfig{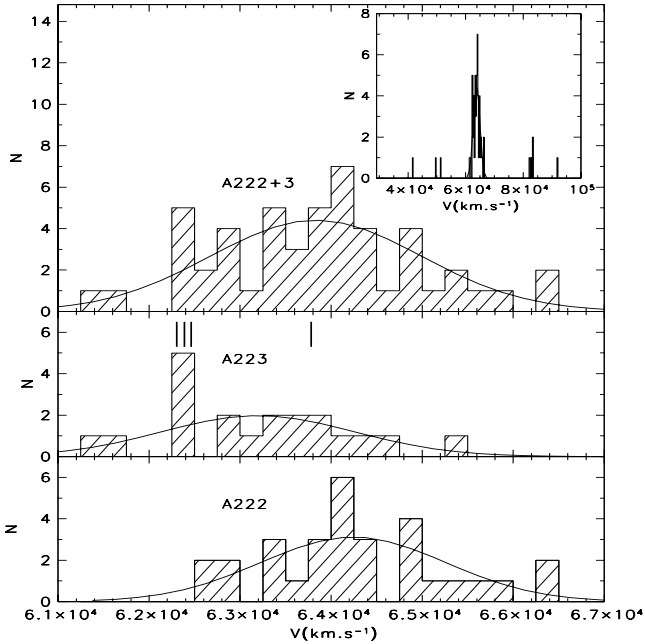}
\caption[]{The radial velocity distribution for the
southern (lower panel) and northern (middle panel) subsamples of
galaxies. The vertical lines in the middle panel display the velocities
of the 4 galaxies belonging to bridge connecting the two main
concentrations. The upper panel shows the velocity distribution of the
sample of both clusters taken together. The continuous lines display
Gaussians curves with means and dispersions values as given in the
text. The inset shows the velocity distribution for the whole sample of
measured galaxies.}
\end{figure}

The lower panels of Figure 5 display the separate velocity distribution
for the southern subsample (30 galaxies), which corresponds to A222,
and for the northern one (20 galaxies), corresponding to A223 (see
Figure 2). The normality tests do not reject Gaussian parent
populations for any of these subsamples. The A222 galaxies have
slightly higher velocities than those of A223: mean velocities are
$V_{bi} = 64242 \pm 194$ km s$^{-1}$, for A222, and $V_{bi} = 63197 \pm
266 $ km s$^{-1}$ for A223.  However, if we remove 4 galaxies belonging
to the bridge connecting the two clusters (see Figure 2), the mean
velocity of A223 increases to $V_{bi} = 63348 \pm 295 $ km s$^{-1}$,
slightly reducing the significance of the velocities difference.
Note that 3 of these galaxies locate at the low velocity tail of
A223, as displayed in Figure~5. The velocity dispersion of the two
clusters are about the same: $\sigma_{bi} = 1013 \pm 150$ km s$^{-1}$
for A222 and $\sigma_{bi} = 1058 \pm 160 $ km s$^{-1}$ for A223 ($1123
\pm 191$ when the bridge galaxies are removed).

Table 2 gives mass and mass-luminosities ratio estimates for each
cluster separately, obtained with the virial and projected mass
estimators given by Heisler et al. (1985), for the case where all the
cluster mass is supposed to be contained in the galaxies, and by
Bahcall \& Tremaine (1981), under the hypothesis that galaxies are
test particles orbiting in a dark mass spherical potential.  
Total $j$ luminosities were
estimated from the BOW83 catalog, which is complete up to $j = 22$.

\begin{table*}
\caption[]{Mass and M/L estimates}
\vspace{0.3cm}
{\centering 
\begin{tabular}{ccccclrccc}
\hline 

\multicolumn{1}{c}{(1)} &
\multicolumn{1}{c}{(2)} &
\multicolumn{1}{c}{(3)} &
\multicolumn{1}{c}{(4)} &
\multicolumn{1}{c}{(5)} &
\multicolumn{2}{c}{(6)} &
\multicolumn{2}{c}{(7)} &
\multicolumn{1}{c}{(8)} \\
\multicolumn{1}{c}{Cluster} &
\multicolumn{1}{c}{R} &
\multicolumn{1}{c}{L$_j$} &
\multicolumn{1}{c}{N$_L$} &
\multicolumn{1}{c}{N$_v$} &
\multicolumn{2}{c}{Mass} &
\multicolumn{2}{c}{M/L} &
\multicolumn{1}{c}{Notes} \\
\multicolumn{1}{c}{} &
\multicolumn{1}{c}{} &
\multicolumn{1}{c}{}&
\multicolumn{1}{c}{}&
\multicolumn{1}{c}{}&
\multicolumn{1}{c}{Virial}&
\multicolumn{1}{c}{Proj.}&
\multicolumn{1}{c}{Virial}&
\multicolumn{1}{c}{Proj.}&
\multicolumn{1}{c}{}\\
\hline 
\hline \\
A222 & 200'' & 1.15 & 107 & 27 & 7.55 $\pm$ 2.56 & 10.05 $\pm$ 3.39 & 658  & 876  & a \\
     &       &      &     &    & 2.85            & 5.22  $\pm$ 1.29 & 248  & 450  & b \\
     &       &      &     &    &                 &                  &      &      &   \\
     & 350'' & 1.46 & 140 & 30 & 8.08 $\pm$ 2.60 &10.25  $\pm$ 3.27 & 553  & 701  & a \\
     &       &      &     &    & 2.91            & 5.30  $\pm$ 1.24 & 199  & 363  & b \\
     &       &      &     &    &                 &                  &      &      &   \\
     &       &      &     &    &                 &                  &      &      &   \\
A223 & 200'' & 1.13 &  84 & 10 & 6.69 $\pm$ 3.73 &12.69  $\pm$ 7.02 & 686  &11230 & a \\
     &       &      &     &    & 3.84            & 7.05  $\pm$ 2.97 & 340  & 624  & b \\
     &       &      &     &    &                 &                  &      &      &   \\
     & 350'' & 1.49 & 124 & 16 &10.17 $\pm$ 4.47 &19.62  $\pm$ 8.58 & 684  &13190 & a \\
     &       &      &     &    & 5.50            &10.46  $\pm$ 3.41 & 370  & 731  & b \\
     &       &      &     &    &                 &                  &      &      &   \\
     &       &      &     &    &                 &                  &      &      &   \\
A520 & 200'' & 2.57 & 166 & 16 &12.33 $\pm$ 5.42 &16.74  $\pm$ 7.33 & 481  & 625  & a  \\
     &       &      &     &    & 6.71            & 8.93  $\pm$ 2.91 & 261  & 348  & b  \\
     &       &      &     &    &                 &                  &      &      &   \\
     & 350'' & 3.78 & 234 & 20 &15.73 $\pm$ 6.19 &20.91  $\pm$ 8.18 & 428  & 569  & a  \\
     &       &      &     &    & 8.01            &11.01  $\pm$ 3.19 & 218  & 300  & b  \\
     &       &      &     &    &                 &                  &      &      &   \\
\hline 
\end{tabular}\par}
\smallskip
{ \small column(1): Cluster Name, column(2): radius in arcsec., column(3): total luminosity with 
$j$ $<$ 22 in $ 10^{12}h^{-2}_{50}L_{\odot}$, 
column(4): number of galaxies with j $<$ 22, column(5): number of galaxies with measured velocity, 
column(6): mass of the cluster in $ 10^{14}h^{-1}_{50}M_{\odot}$,
column(7): mass-luminosity ratio in solar units,
column(8): mass estimators: (a) self-gravitating system, (b) test particle system.}
\end{table*}

\subsection{The cluster A520}

The analysis of A520 data proceeded in the same lines as that of the
A222+A223 system. Figure 6 displays the significance map of projected
densities of the DSS galaxies in the field of A520. This figure also
displays the positions of 21 galaxies having measured radial velocities
and belonging to the cluster, as discussed below. As for the case of
A222/3, here also we can see that the main concentration has two extensions,
possibly due to infalling clumps of galaxies. 

In Figure 7 we display the X-ray isophotes of a wavelet reconstruction
of the ROSAT/HRI image of A520, superposed to the significance map of
the projected density of galaxies (dashed lines). As before, this map
was constructed by taken galaxies from the BOW83 catalog, showing that
the cluster may be much more complex than it could be noticed from the
DSS map, although the reality of the substructures shown here cannot be
assigned in view of the paucity of radial velocity data.

As it can be seen from this figure, although the main X-ray emission
roughly follows the projected density of galaxies, it seems dislocated
relatively to main concentration of A520, located near the center
of this field. The peak X-ray emissivity comes from a very compact
region which may be consistent with a point-like source, almost
coincident with a ``blue'' galaxy ($j-f_{57} = 1.22$), originally
assigned by Sandage, Kristian \& Westphal (1976) as one of the brightest 
A520 members
(it ranks a 7th place in $j$ magnitudes but only 19th in $f_{57}$
magnitudes). In fact, as displayed in Figure 7, the $f_{57}$ brightest
members of A520 do not seem to belong to any of the main galactic
clumps observed, a situation which is similar to that already noticed
in the case of A222. This means that most of the clumps are constituted
by the faint galaxy population, not present in the DSS sample. We may
conclude that, unless we are facing a serious case of background
contamination, the A520 cluster, as for A222, may be an example of a
dynamically young system where clumps of galaxies are
still in phase of collapsing
on its dark matter gravitational well, probably located at the mean
center of the X-ray emission region seen in Figure 7. Unfortunately
there is no X-ray spectra available for A520 (as also for A222/3),
what hinders a more detailed diagnostic of the evolutive dynamical
stage of the cluster.

\begin{figure}
\psfig{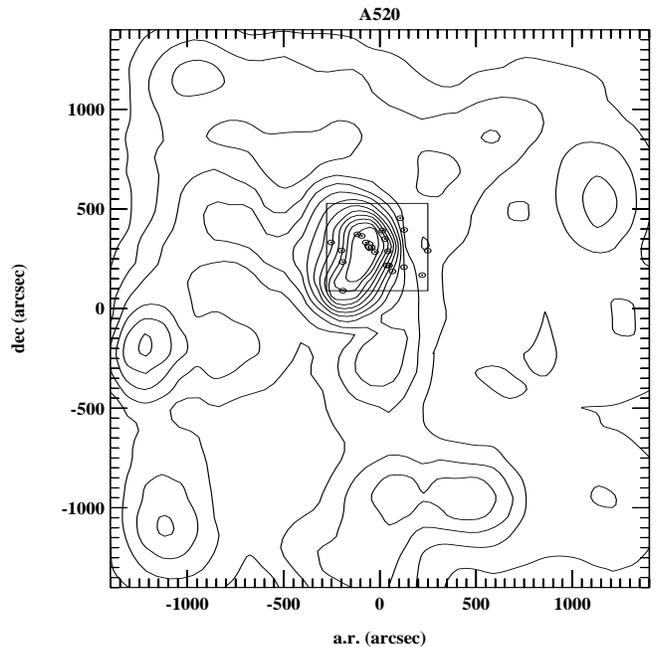}
\caption[]{Projected density map ({\it
significance map}) of the galaxies in the field of A520 
(centered at $\alpha_{2000} = 04^h 54^m 18^s$, 
$\delta_{2000} = +02^\circ 52^\prime 00^{\prime \prime}$). 
The symbols have the same meaning as
in Figure 2.}
\end{figure}

\begin{figure}
\psfig{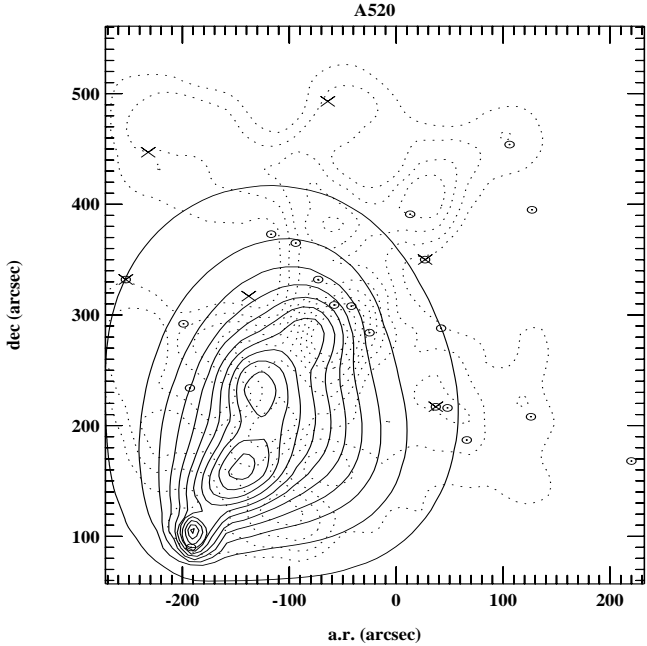}
\caption[]{Like Figure 3a but for A520. The 6
brightest members of the cluster are marked with crosses.}
\end{figure}

Our sample of spectroscopically measured objects in the field of A520
(Table 1) has 28 galaxies, with 25 coming from the observations
reported here and 3 others from the literature (Sandage, Kristian \&
Westphal 1976; Newberry, Kirshner \& Boroson 1987). 
 
Figure~8 shows the wedge velocity diagram in right ascension and
declination for A520.

\begin{figure*}
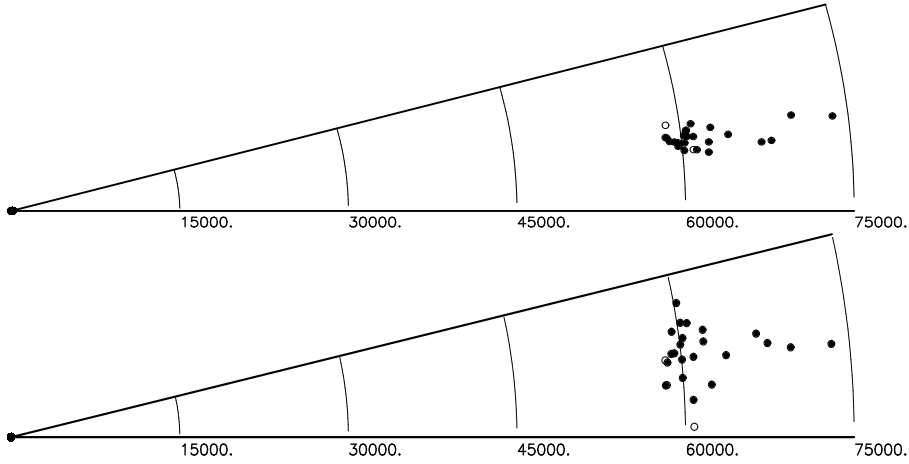

\psfig{figure=FIG8a.epsi,height=3cm,width=12cm,angle=-90}
\psfig{figure=FIG8b.epsi,height=3cm,width=12cm,angle=-90}
\caption{Wedge velocity diagram in right
ascension (up), and declination (down) for the measured galaxies in
A520. Filled symbols represent galaxies measured in the present work;
open symbols represent galaxies with velocities from the literature.}
\end{figure*}

The $3\sigma$ clipping of the total radial velocity distribution
leaves 21 galaxies kinematically linked to the cluster. This sample is
consistent with normality under all the statistical tests included in
the ROSTAT routine.  For comparison, we applied the same tests to a
sample including the less discrepant galaxy, in velocity space,
excluded by the $3\sigma$ clipping.  Although the tests fail to reject
the normality for this sample, it resulted skewed towards higher
velocities, as indicated by the Asymmetry Index obtained: AI = 0.85.

Figure 9 displays the velocity distribution of the 21 retained
galaxies as well as that of the whole sample (inset). The mean velocity
for this sample  is $V_{bi} = 60127 \pm 284$ km s$^{-1}$, with
dispersion $\sigma_{bi} = 1250 \pm 189$ km s$^{-1}$, placing the
cluster at redshift $ z = 0.20056 $. The values for the mass and
mass-luminosity ratio, calculated under the same hypothesis as for A222
and A223, are given in Table 2.

\begin{figure}
\psfig{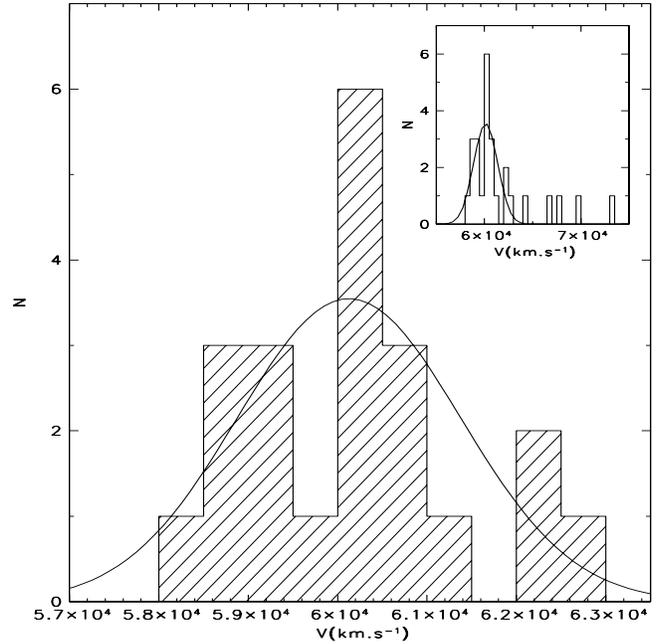}
\caption[]{The radial velocity distribution for the
A520 sample of galaxies. The continuous curve shows the Gaussian
distribution corresponding to the mean velocity and dispersion quoted in
the text (normalized to the sample size and range). The inset shows
the velocity distribution for the whole sample of measured galaxies.}
\end{figure}

\section{Spectral Classification}

Spectral classification has been performed through a Principal Component Analysis
(PCA) of the spectra. This technique makes use of all information contained in the
spectra (except in the emission lines; see below) and, in this sense, it can provide
a classification scheme more powerful than those based on the amplitude of individual
absorption lines.
Here we apply the method to a sample of 51 CFHT spectra of 
galaxies that are probably members of the clusters A222, A223 and A520 (section 3)
to obtain spectral types.  
The point to be stressed is that the spectra of
normal galaxies form a sequence- the spectral sequence- in the spectral space 
spanned by the $M$-dimensional vectors that contain the spectra, each vector being
the flux of a galaxy (or a scaled version of it) sampled at $M$ wavelengths
(Sodr\'e \& Cuevas 1994, 1997; Connolly \etal 1995; Folkes \etal 1996). 
The spectral sequence correlates well with the Hubble
morphological sequence, and we define the ``spectral type'' (hereafter ST) of a
galaxy from its position along the spectral sequence. Following Sodr\'e \& Cuevas (1997),
we associate the spectral type ST of a galaxy with its value for the first principal
component. Note that, since we are working with uncalibrated spectra, only a 
relative classification is possible, that is, we are only able to know whether a 
galaxy has an earlier or later spectral type than the others in the 
sequence (Cuevas, Sodr\'e \& Quintana 2000). 

We have jointly analyzed the spectra of galaxies in the three clusters
because the differences in their redshifts are small and the observed spectra sample
essentially the same rest-frame wavelength interval. PCA was applied to a
pre-processed version of 51 CFHT uncalibrated galaxy spectra (33 were spectra of
galaxies of A222 and A223 and 18 of A520). 
Firstly, the spectra were shifted to the
rest frame and re-sampled in the wavelength interval from 3440 \AA \( \:  \)to 
5730 \AA, in equal-width bins of 2 \AA.  Secondly, we removed from the analysis 8
regions of \( \sim  \)40 \AA\( \:  \)each centered at the wavelengths of {[}OII{]}
\( \lambda  \)3727, NeIII \( \lambda  \)3869, H\( \delta \) \( \lambda  \)4102, 
H\( \gamma  \) \( \lambda  \)4340, HeII \(\lambda  \)4686, H\( \beta  \)
\( \lambda  \)4861, {[}OIII{]} \(\lambda  \)4959 and {[}OIII{]} \( \lambda  \)5007.
This was done in order to avoid the inclusion of emission lines in the analysis,
which increases the dispersion of the spectra in the principal plane (mainly due to
an increase in the second principal component). The spectra, now sampled at 
$M=980$ wavelength intervals, were then normalized to the same mean flux 
($\sum_\lambda f_\lambda= 1$). Finally, we subtracted the mean spectrum from the
spectrum of each galaxy and use the PCA to obtain the principal components. This
procedure is equivalent to the conventional PCA on the covariance matrix (that is,
the basis vectors are the eigenvectors of a covariance matrix).
 
Figure~10 shows the projection of the spectra of the 51 galaxies of the three 
clusters on to the plane defined by the first two principal components. They contain 
only 24\% of the total variance, mainly due to the low signal-to-noise ratio of
several spectra (the median S/N is \( \sim \)5.8 in the interval between 4500 \AA 
\( \:  \)to 5000 \AA). Indeed, in this figure different symbols correspond to
different  signal-to-noise  intervals (see the figure caption), and the scatter 
in the second principal component seems to increase as the signal-to-noise ratio 
decreases. 
On the other side, numerical simulations indicate that the noise does not
introduce any significant bias in the spectral classification 
(Sodr\'e \& Cuevas 1997). Note that, in this
figure, early-type galaxies are at the left side, and increasing values of ST 
(or, equivalently of the first principal component) correspond to later-type 
galaxies. 

The low variance accounted for by the two first principal components may rise 
doubts about whether we are indeed measuring meaningful spectral types through the
first principal component. A possible approach is to compare our classification 
with the spectral classification present in Newberry, Kirshner \& Boroson  
(1988). These authors 
classified a few galaxies in A222, A223, and A520 in red or blue accordingly to
their colors and position in a color-magnitude diagram, or from the
strength of some absorption features present in the spectra. Unfortunately
we have only four cluster galaxies in common with Newberry, Kirshner 
\& Boroson (1988).
Nevertheless our results are encouraging, because the three galaxies classified
as red by Newberry, Kirshner \& Boroson (1988) have spectral types equal to 
or smaller than 
$\sim$1.5, while the only galaxy classified as blue by them has a spectral type
of~2.46. Additionally, as we will show, with these spectral types we are able to
recover both the morphological and kinematical segregations for the galaxies in these
clusters, indicating that our spectral types are indeed carrying useful morphological
information.

\begin{figure}
\psfig{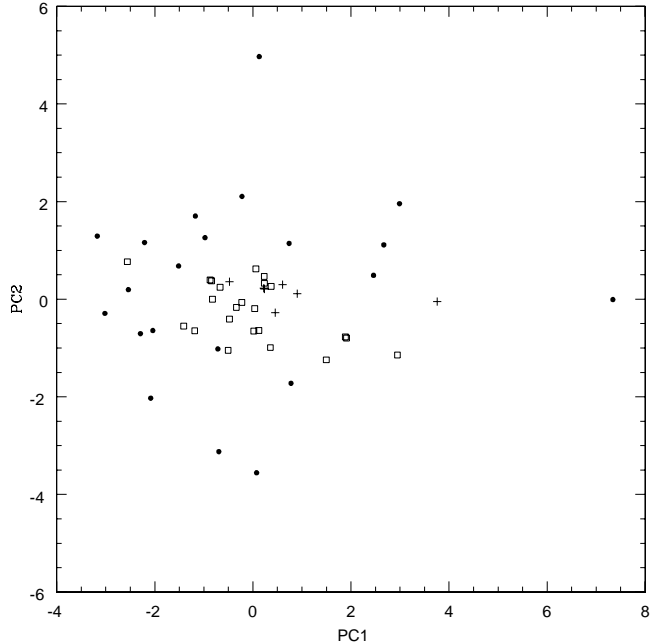}
\caption[]{Projection of the spectra of A222, A223 and
A520 onto the first two principal components of the galaxies. Different symbols
correspond to different signal-to-noise ratios, computed in the wavelength interval
between 4500\AA~ and 5000\AA. Filled circles: $2.9\le$ S/N $<5.5$; squares
$5.5\le$ S/N $<8.0$; crosses: S/N $\ge 8.0$.}
\end{figure}

It is worth emphasizing that we base our spectral classification only on the 
properties of the stellar populations that are contained in the continuum and
absorption lines, and that the emission lines enter in no way in the classification
scheme. It is important to point out, however,
that the emission lines of normal galaxies do correlate with spectral types; see
Sodr\'e \& Stasi\'nska (1999) for a detailed discussion of this subject.

\section{Morphological and Kinematical Segregation}

Now we use the spectral types of the galaxies to study whether the morphology-density 
relation (Dressler 1980) is present in these clusters. We have computed the
projected local density from the 6~nearest (projected) neighbors of each of the
galaxies in our spectroscopic sample with the estimator (Casertano \& Hut 1985):
\[
\rho _{proj}=\frac{5}{\pi r^{2}_{6}}\]
\noindent where \( r_{6} \) is the projected distance of the \( 6^{th} \)
nearest galaxy. We have used the catalogue obtained from
DSS (see section 3) to estimate the local density. 
Figure~11 shows the logarithm
of the projected local density normalized by the median density of each cluster
versus the spectral type. 

\begin{figure}
\psfig{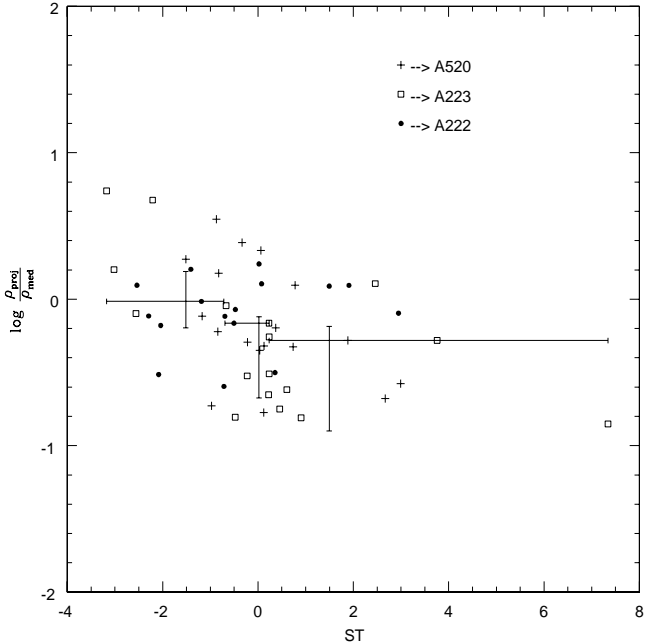}
\caption[]{Projected local density normalized by the median density of each 
cluster versus the spectral type. The points with error bars are the median
values taken in bins of equal galaxy number. The vertical error bars correspond 
to the quartiles of the distribution, whereas the horizontal error bars indicate
the interval of ST associated to each bin. Different symbols correspond to data of
different clusters.}
\end{figure}

This figure shows that, for A222, A223 and A520, early-type galaxies tend to be
located in denser regions than late-type galaxies, indicating that the 
morphology-density relation (Dressler 1980), as inferred using spectral types, 
was already established in clusters
at $z\sim 0.2$. The correlation shown in Figure~11 is significant: the Spearman
rank-order correlation coefficient is \( r_s \) = -0.41 and the two-sided 
significance level of its deviation from zero is \( p \) = 0.002.

Nearby clusters also present a ``kinematical segregation'': the
velocity dispersion of early-type galaxies is lower than those of
late-type galaxies (Sodr\'e et al. 1989). This may be an evidence that
late-type galaxies have arrived recently in the cluster and are not yet
virialized, while the early-type galaxies constitute a relaxed systems
with a low velocity dispersion. We present in
Figure~12, as a function of spectral type, the absolute value of galaxy
velocities relative to the clusters mean velocity, normalized by the
velocity dispersion of each cluster.  The points with error bars in this
figure are the median values taken in bins of equal galaxy number;  the
vertical error bars correspond to the quartiles of the distribution, while
the horizontal ones indicate the interval of ST corresponding to each bin. 
The data in Figure~12 indicate that early-type galaxies tend to have lower
relative velocities than galaxies of later types; the Spearman rank-order
correlation coefficient \( r_s \) is now 0.39 and the two-sided significance
level of its deviation from zero is 0.005. Hence, these clusters seem to 
present the same kind of kinematical segregation detected in low redshift
galaxy clusters.

\begin{figure}
\psfig{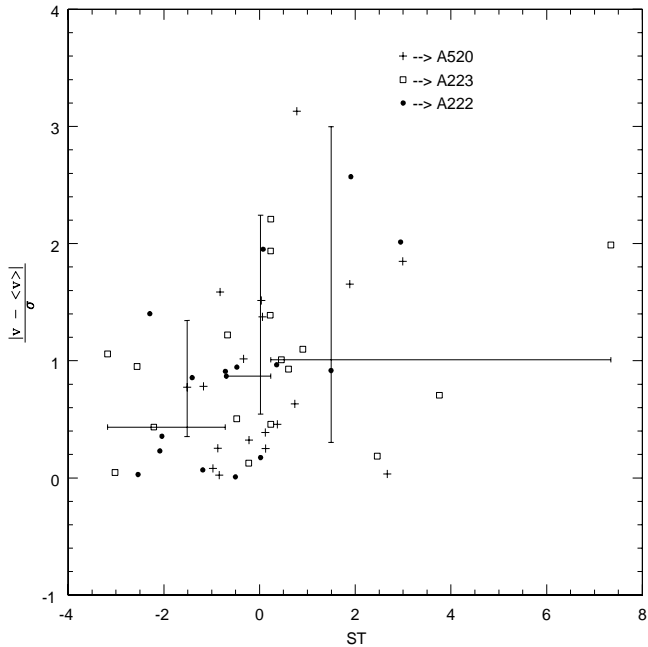}
\caption[]{Velocity corrected to the cluster mean velocity and normalized by
the velocity dispersion of each system versus the spectral type. The three
clusters are plotted together, with different symbols corresponding to different
clusters. The errors bars have the same meaning as in Figure~11.}
\end{figure}

\section{Summary}
We have presented here an analysis of three medium redshift clusters,
A222, A223, and A520, based on new observations of radial velocities in
the field of these clusters.  Through observations made at the
Canada-France-Hawaii Telescope, the European Southern Observatory, and
the Pic du Midi Observatory, we obtained a set of 78 new redshifts, 71
of them corresponding to galaxies members of these clusters.

From these observations and velocities and X-ray data from the
literature we concluded that
A222 and A223 have similar radial velocities and velocity dispersions,
and will probably merge in the future, as already suggested by Sandage,
Kristian \& Westphal (1976).  
A520 also seems to be undergoing strong dynamical
evolution, since its cD galaxy is not located at the center of the
galaxy distribution (that is also coincident with the X-ray emission).

We have used spectra taken at CFHT to obtain, through a Principal
Component Analysis, spectral types for a subset of 51 galaxies in these
clusters. We have shown that galaxies of ``early'' spectral types tend
to be found in regions with densities larger than that where ``late''
spectral type galaxies are found, suggesting that the morphology -
density relation was already established at $z\sim 0.2$.  We have also
found that galaxies with ``early'' spectral types tend to have lower
velocity dispersions when compared with ``late'' spectral type
galaxies, evidencing that the kinematical segregation too was already
established at intermediate redshifts.  These results are interesting
because, despite the fact that these clusters are probably in a stage
of strong evolution, they already show features that are expected for
relaxed structures, as is the case of the segregations mentioned
above.

\acknowledgements{We thank Christian Vanderriest for his collaboration
in the CFHT observations and the CFHT, ESO and TBL staff.
BTL, HC, HVC, and LSJ have benefited from the support
provided by FAPESP, CNPq and PRONEX/FINEP to their work. 
We also thank an anonymous referee for useful comments that allowed us
to improve the paper.}

{\bf References}

Bahcall J.N., Tremaine S., 1981, ApJ, 244,805.

Beers T.C., Flynn K., Gebhart K., 1990, AJ, 100, 32.

Bertin E., Arnouts S., 1996, A\&AS, 117, 393.

Bird C.M., Beers T.C., 1993, AJ, 105, 1596.

Biviano A., Durret F., Gerbal D., le Fevre O., Lobo C.,
Mazure A., Slezak E., 1996, A\&A, 311, 95.

Bromley B.C., Press W.H., Lin H., Kirshner R.P., 1998,
ApJ, 505, 25.

Butcher H., Oemler A., Wells D.C., 1983, ApJS, 52, 183 (BOW83).

Casertano S., Hut P., 1985, ApJ, 298, 80.

Cole S., 1989, PhD Thesis.

Connolly A.J., Szalay A.S., Bershady M.A., Kinney A.L.,
Calzetti D., 1995, AJ, 110, 1071.

Cuevas H., Sodr\'e L., Quintana H., 2000, in preparation.

Danese L., De Zotti G., di Tullio G., 1980, A\&A, 82, 322.

Dressler A., 1980, ApJS, 42, 565.

Folkes S.R., Lahav O., Maddox S.J., 1996, MNRAS, 283, 651.

Galaz G., de Lapparent V., 1998, A\&A, 332, 459.

Hartigan J.A., Hartigan P.M., 1985, Annals of Stat., 13, 70.

Heisler J., Tremaine S., Bahcall J.N., 1985, ApJ, 298, 8.

Kurtz M.J., Mink D.J., Wyatt W.F., Fabricant D.G., Torres 
G., Kriss G.A., Tonry J.L., 1991, ASP Conf. Ser., 25, 432.

Lea S.M., Henry J.P., 1988, ApJ, 332, 81.

Le F\`evre O., Crampton C., Lilly S.J., Hammer F., Tresse L., 1995,
ApJ, 455, 60.

Mink D.J., Wyatt W.F., 1995, ASP Conf. Ser., 77, 496.

Newberry M.V., Kirshner R.P., Boroson T.A., 1988, ApJ, 335, 629.

Ribeiro A.L.B., de Carvalho R.R., Capelato H.V., Zepf S.E., 1998, 
ApJ, 497, 72.

Ru\'e F., Bijaoui A., 1997, Experim. Astron., 7, 129.

Sandage A., Kristian J., Westphal J.A., 1976, ApJ, 205, 688.

Sodr\'e L., Capelato H.V., Steiner J.E., Mazure A., 1989, AJ, 97, 1279.

Sodr\'e L., Cuevas H., 1994, Vistas in Astronomy, 38, 287.

Sodr\'e L., Cuevas H., 1997, MNRAS, 287, 137.

Sodr\'e L., Stasi\'nska G., 1999, A\&A, 345, 391.

Struble M.F., Rood H.J., 1987, ApJS, 63, 543.

Tonry J., Davis M., 1979, AJ, 84, 1511.

West M.J., Oemler A.,. Dekel A., 1989, ApJ, 346, 539.

Yahil, A., Vidal, N.V., 1977, ApJ, 214, 347.
 
\end{document}